\documentclass[bst/sn-nature]{sn-jnl}

\usepackage[T1]{fontenc}

\usepackage{type1cm}        
                            
\usepackage{makeidx}         
\usepackage{graphicx}        
\usepackage{multicol}        
\usepackage[bottom]{footmisc}

\usepackage{newtxtext}       %
\usepackage[varvw]{newtxmath}       

\usepackage{svg}
\usepackage{float}
\usepackage{array}
\usepackage{adjustbox}
\usepackage{caption}
\usepackage{booktabs}
\usepackage{color, colortbl}
\usepackage{xcolor, soul}
\usepackage{tabularx}
\sethlcolor{yellow}
\usepackage{subcaption}
\usepackage{enumerate}

\usepackage{pdflscape}
\usepackage{multirow}
\usepackage{mdframed}

\hypersetup{
    colorlinks      = {true},
    linkcolor       = {blue}, 
    linkbordercolor = {white},
    citecolor       = {blue},
    urlcolor        = {blue}
}
      
\usepackage[super]{nth}

\usepackage{comment}
\usepackage{todonotes}

\usepackage{doi}
\usepackage{orcidlink}
\usepackage{listings}
\usepackage{tcolorbox}
\usepackage{longtable}

\definecolor{resetting_color}{HTML}{09777A}
\definecolor{back_color}{HTML}{DCEEEF}
\tcbuselibrary{breakable}
\usepackage{placeins}

\newfloat{colorbox}{htbp}{lob}
\floatname{colorbox}{Box}
\begin{document}

\title{Smart ETL and LLM-based contents classification:}
\subtitle{the European Smart Tourism Tools Observatory experience}

    
\author*[1]{\fnm{Diogo} \sur{Cosme}\orcidlink{0009-0001-1245-286X}}
\email{dfmce@iscte-iul.pt}

\author*[2]{\fnm{António} \sur{Galvão}\orcidlink{0000-0002-6566-9114}}
\email{amg13172@campus.fct.unl.pt}
\equalcont{These authors contributed equally to this work.}

\author*[1]{\fnm{Fernando} \sur{Brito e Abreu}\orcidlink{0000-0002-9086-4122}}
\email{fba@iscte-iul.pt}
\equalcont{These authors contributed equally to this work.}

\affil[1]{\orgdiv{ISTAR}, \orgname{ISTA, Instituto Universitário de Lisboa (Iscte-IUL)}, \orgaddress{\street{Av. das Forças Armadas, 40}, \city{Lisboa}, \postcode{1649-026}, \country{Portugal}}}

\affil[2]{\orgdiv{CENSE}, \orgname{School of Science and Technology, NOVA University Lisbon}, \orgaddress{\street{Campus da Caparica}, \city{Caparica}, \postcode{2829-516}, \country{Portugal}}}

\abstract{\textbf{Purpose:}
Our research project focuses on improving the content update of the online European Smart Tourism Tools (STTs) Observatory by incorporating and categorizing STTs. The categorization is based on their taxonomy, and it facilitates the end user's search process. The use of a Smart ETL (Extract, Transform, and Load) process, where \emph{Smart} indicates the use of Artificial Intelligence (AI), is central to this endeavor.

\textbf{Methods:}
The contents describing STTs are derived from PDF catalogs, where PDF-scraping techniques extract QR codes, images, links, and text information. Duplicate STTs between the catalogs are removed, and the remaining ones are classified based on their text information using Large Language Models (LLMs). Finally, the data is transformed to comply with the Dublin Core metadata structure (the observatory's metadata structure), chosen for its wide acceptance and flexibility.

\textbf{Results:}
The Smart ETL process to import STTs to the observatory combines PDF-scraping techniques with LLMs for text content-based classification. Our preliminary results have demonstrated the potential of LLMs for text content-based classification.

\textbf{Conclusion:}
The proposed approach's feasibility is a step towards efficient content-based classification, not only in Smart Tourism but also adaptable to other fields. Future work will mainly focus on refining this classification process.
}


%


\maketitle


\section{Introduction}
\label{sec:intro}


The research presented in this paper aligns with the broader goals of advancing knowledge systems and enhancing information retrieval by leveraging large language models (LLMs) to classify Smart Tourism Tools (STTs) within an automated ETL process for extracting, transforming, and classifying data from complex, unstructured sources like PDFs. 

The use of LLMs in information retrieval, particularly for content classification, is a recent research topic where we could not find a systematic literature review (SLR), although many primary studies have already been published. Due to its importance, we have produced such SLR on this topic (\cite{Cosme2024a}, supplemented by \cite{Cosme2024b}). Therefore, instead of including a dedicated section on related work in this article, we invite the reader to consult the aforementioned SLR for a comprehensive review of this area.

\begin{figure}[h]
    \centering
    \includegraphics[width=.9\textwidth]{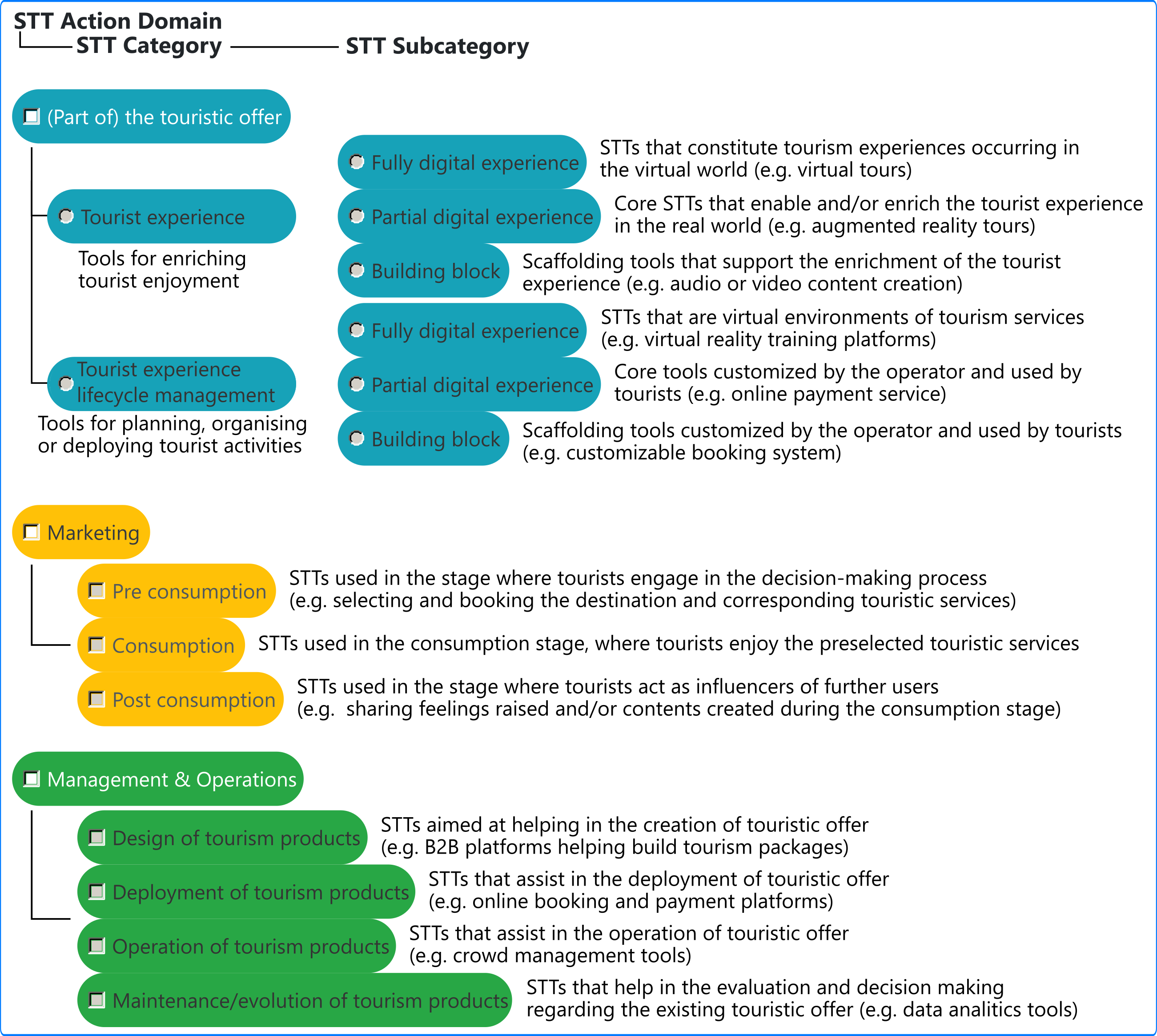}
    \caption{STT Taxonomy}
    \label{fig:stts_domains_and_types}
\end{figure}

Smart Tourism (ST) is a usual designation for ICT-based innovation in the tourism sector. The corresponding ICT tools are dubbed Smart Tourism Tools (STTs). A literature review on this topic \cite{Galvao2024} concluded that \emph{``Despite all the hype around ST, there is a lack of consensus on the definition of ST''} and proposed a definition of STTs, drawing from the insights of 330 worldwide tourism experts: \emph{"seamlessly interconnected digital tools designed to benefit all stakeholders in the tourism industry, with a special focus on the tourist and the destination, that aim at sustainable development"}. A taxonomy was also proposed there to aid in organizing these STTs, based on products, services, and applications ``made in Europe''. The taxonomy is divided into three application domains: \textit{(Part of) the Touristic Offer}, \textit{Marketing}, and \textit{Management \& Operations}, each further subdivided into STT categories, and in some cases subcategories, as represented in Figure~\ref{fig:stts_domains_and_types}.

The work underlying this paper was developed in the framework of the \href{https://www.resetting.eu/}{RESETTING} project\footnote{\textbf{R}elaunching \textbf{E}uropean \textbf{S}mart and sustainabl\textbf{E} \textbf{T}ourism models \textbf{T}hrough digitalization and \textbf{IN}novative technolo\textbf{G}ies}, funded by the \href{https://single-market-economy.ec.europa.eu/smes/cosme_en}{European Union's COSME program}. Its main objective was to support the transformation of European tourism businesses towards more resilient, circular, and sustainable operating models by testing and implementing cutting-edge, digitally-driven solutions that reduce unnecessary burdens, enhance the quality of the travel experience, support the decarbonization of the tourism sector, and promote more equitable economic growth that benefits both SMEs (Small and Medium Enterprises) and destination residents. RESETTING aimed to help SMEs overcome the challenge of not having the resources and skills to keep up with technological developments and, as a result, presented innovative solutions. Participating business associations, universities, and public sector organizations throughout the project sought to stimulate business innovation in European tourism SMEs.

    
The tourism industry's shift towards digitalization and innovation requires STTs. However, there is a digital divide between large enterprises and SMEs in terms of their stake in the tourism industry \cite{Minghetti2010,Reverte2020}, and their ability to capitalize on the opportunity of a digital transformation of their core business due to financial and technical limitations, as well as an unawareness of existing STT. The RESETTING project proposed to address the unawareness limitation by developing the \href{https://sttobservatory.resetting.eu}{European STT Observatory}. It aimed to provide a comprehensive and up-to-date overview of STT provision in Europe. However, the goal goes beyond creating and populating the Observatory with STTs. The aim is also to automatically categorize them based on the STT taxonomy, making it easier for users to find exactly what they need.

\label{sec:expected_results}
The main result of our work will be the smart ETL approach used to load the European STT Observatory with classified content. This will allow its users to perform more guided and efficient searches and find the type of STT they are looking for more easily. Ultimately, this approach will allow tourism SMEs to effectively and efficiently identify the STTs available on the European market (or just in their region) to boost innovative and sustainable activities.



This article is structured as follows: section \ref{sec:background} covers the necessary background context, sections \ref{sec:smart_extract}, \ref{sec:smart_transformation}, and \ref{sec:load_phase} detail the developed and implemented methodology, section \ref{sec:results} discusses the main results and identifies its main limitations, section \ref{sec:verification_and_validation} focuses on the verification and validation of the proposal, section \ref{sec:conclusions} draws some conclusions and, finally, future work is outlined on section \ref{sec:future_work}.

\section{Background}
\label{sec:background}

\subsection{The observatory}

As the backbone for our observatory, we chose the \href{https://omeka.net/}{Omeka Net} platform, a hosted version of the open source \href{https://github.com/omeka}{Omeka platform}\footnote{Omeka is a web publishing platform for sharing digital collections and creating media-rich online exhibits}, because of its stability and widespread use. It also includes an API that allows us to automatically add STTs to the Observatory. It works through HTTP requests with the data field in JSON format. Tasks related to inserting, updating and deleting content require an API key. Omeka allows us to define items, collections, and exhibits. Items are the most basic element and it is possible to upload files to them. In our case, an item represents an STT. A collection is like a folder that groups related items, and an exhibit is a way to combine items in a narrative text. While an item can be in multiple exhibits, it can only be in one collection.

  
\subsection{STTs catalogs}
\label{sec:catalogs}

The current observatory content is extracted from two types of catalogs:
\begin{itemize}
    \item \textbf{Catalogs of STTs:} This category includes four catalogs, including the 2022 and 2023 versions of the \cite{SEGITTUR2023} catalog, produced by \href{https://www.segittur.es/}{SEGITTUR - Sociedad Mercantil Estatal para la Gestión de la Innovación y las Tecnologías Turísticas} in Spain. From now on, they will be identified as \href{https://sttobservatory.omeka.net/collections/show/1}{\textit{SEGITTUR 2022}} and \href{https://sttobservatory.omeka.net/collections/show/9}{\textit{SEGITTUR 2023}}, respectively. The other two catalogs are the first and second versions of the \cite{ADESTIC2023}, produced by \href{https://adestic.org/es/}{ADESTIC - Clúster de Empresas Innovadoras para el Turismo de la Comunitat Valenciana} also in Spain, which will be referred to as \href{https://sttobservatory.omeka.net/collections/show/8}{\textit{ADESTIC V1}} and \href{https://sttobservatory.omeka.net/collections/show/7}{\textit{ADESTIC V2}}, respectively. These were used as data sources for the STT classification.
    
    \item \textbf{Catalogs of ST Practices:} Rather than listing various STTs, these catalogs detail services and initiatives implemented using STTs, categorized as ST practices. The versions used were 2022 and 2023 of the \cite{Scholz2023}, commissioned by the European Commission, which will be referred to as \href{https://sttobservatory.omeka.net/collections/show/3}{\textit{EU 2022}} and \href{https://sttobservatory.omeka.net/collections/show/4}{\textit{EU 2023}}, respectively. Since these catalogs do not describe STTs but rather their use, their content was not used in the classification task.
    
\end{itemize}

The \textit{SEGITTUR} and \textit{ADESTIC} catalogs contain only STTs developed by Spanish companies. In addition, the \textit{ADESTIC} catalogs are in Spanish, so the extracted content must be translated into English. As for the European Commission catalogs, they only list services and initiatives implemented in Europe that use STTs. Next, we will examine in detail the structure and types of content present in each catalog.

The \textit{SEGITTUR} and \textit{ADESTIC} catalogs share a similar structure. Both include an iconographic glossary featuring various graphical icons, each associated with a specific label, representing the types of solutions designated by the respective catalog producers. For each STT, the associated icons are extracted and matched against those in the glossary to obtain the corresponding label. \textit{SEGITTUR} also offers an additional iconographic glossary illustrating the types of destinations suitable for STTs (i.e., Culture and Urban, Nature and Sport, Niche, and Beach). Thus, a process similar to the one described previously is carried out. \textit{ADESTIC} also identifies the types of destinations suitable for STTs but uses a text label instead of an icon.

The relevant elements the Spanish catalogs have in common are presented in Table \ref{tab:spanish_elements}.
\begin{table}[ht]
\caption{Common elements of \textit{ADESTIC} and \textit{SEGITTUR} catalogs}
\label{tab:spanish_elements}
\centering
\fontsize{8pt}{10pt}\selectfont
\renewcommand{\arraystretch}{1.1}
\begin{tabular}{|l|p{9.25cm}|}
\hline
\textbf{Property} & \textbf{Description} \\
\hline
STT's Name & The name of the Smart Tourism Tool.\\
\hline
STT's Description & A textual description of the STT. \\
\hline
Producer's Name & The name of the company that produced the STT. \\
\hline
Producer's Description & A text description of the producer. \\
\hline
Producer's Logo & A graphical logo of the producer. \\
\hline
Producer's URL & The URL of the STT producer.\\
\hline
Type of Destination & In \textit{ADESTIC} catalogs, it appears as a text element; in \textit{SEGITTUR} catalogs, it is represented as an icon. \\
\hline
Type of Solution & This element is an icon in both catalogs. \\
\hline
\end{tabular}
\end{table}

In addition to these elements, there are unique features. \textit{SEGITTUR} catalogs identify the scope of the STT (e.g., governance, technology, sustainability, innovation, or accessibility) with a text label. \textit{ADESTIC} catalogs include the producer's phone number, email address, and QR codes pointing to videos and additional files. The phone prefix may also identify the producer's regional location for landlines.

Concerning European catalogs, the relevant elements are presented in Table \ref{tab:elements_europe}
\begin{table}[ht]
\caption{STT-based initiative properties and their descriptions}
\label{tab:elements_europe}
\centering
\fontsize{8pt}{10pt}\selectfont
\renewcommand{\arraystretch}{1.1}
\begin{tabular}{|l|l|}
\hline
\textbf{Property} & \textbf{Description} \\
\hline
STT-based initiative title & The designation of the STT-based initiative.\\
\hline
STT-based initiative location & The city where each service or initiative was deployed. \\
\hline
STT-based initiative description & A text description of the service or initiative implemented. \\
\hline
STT-based initiative URL & The URL where the application of the STT is reported.\\
\hline
STT-based initiative image(s) & associated image(s), if any. \\
\hline
\end{tabular}

\end{table}

    
\subsection{Observatory ontology}

Each item published in the observatory must match the Dublin Core metadata structure to describe digital or physical resources. The \href{https://www.dublincore.org/specifications/dublin-core/dces/}{Dublin Core}, also known as the Dublin Core Metadata Element Set (DCMES), is a set of fifteen core metadata elements that have been standardized as \href{https://www.iso.org/standard/71339.html}{ISO 15836-1}, \href{https://www.ietf.org/rfc/rfc5013.txt}{IETF RFC 5013}, and \href{https://www.niso.org/publications/ansiniso-z3985-2012-dublin-core-metadata-element-set}{ANSI/NISO Z39.85}. Each element has a Uniform Resource Identifier (URI)\footnote{A URI is a string of characters used to identify an abstract or physical resource. There are two types of URIs: Uniform Resource Locator (URL), to specify the location of a resource; and Uniform Resource Name (URN), to identify a resource by name within a particular namespace.}, and they have been assigned to the \href{http://purl.org/dc/elements/1.1/)}{\emph{dc:} namespace}, i.e., to ensure consistent identification and use across applications and systems, each URI has the prefix \emph{dc:}. Table \ref{tab:dcmes_elements} provides descriptions of some of these elements, with the \emph{dc:} namespace included in their URIs.

These core elements now belong to a wider set of metadata vocabularies and technical specifications maintained by the \href{https://www.dublincore.org/}{Dublin Core Metadata Initiative (DCMI)}, named \href{https://www.dublincore.org/specifications/dublin-core/dcmi-terms/}{DCMI Metadata Terms (DCTERMS)}, and available on the \href{http://purl.org/dc/terms/)}{\emph{dcterms:} namespace}. It was also standardized as \href{https://www.iso.org/standard/71341.html}{ISO 15836-2}. When mentioning Dublin Core, \textit{elements} refers to the DCMES properties, while \textit{terms} refers to the DCTERMS.

Compared to the definition of \textit{elements}, \textit{terms} are more detailed and precise, including the type of term, formal ranges, and classes. To maintain compatibility with existing \href{https://www.w3.org/RDF/}{RDF (Resource Description Framework)}\footnote{RDF is a standard model for data interchange on the Web.} implementations of DCMES, fifteen \textit{terms} with the same names were created in DCTERMS and defined as sub-properties of the corresponding DCMES \textit{elements}. This compatibility is essential for several reasons: it ensures interoperability, allowing different systems and applications to work together seamlessly; it ensures data consistency, allowing data described with DCMES to be understood and processed accurately; it maintains backward compatibility, preventing current implementations from becoming obsolete; and it supports a seamless transition for organizations and developers, allowing them to adopt the new terms without making major changes to their existing RDF-based systems. Table \ref{tab:dcterms_terms} contains the definitions of the \textit{terms} for the corresponding \textit{elements} shown in Table \ref{tab:dcmes_elements}.

\begin{table}[h!]
\vspace{-0.5cm}
\renewcommand{\arraystretch}{1.25}
\centering
\caption{Some descriptions of DCMES elements} 
\label{tab:dcmes_elements}
\begin{tabular}{|p{1.25cm}|p{11cm}|}
\hline
\multicolumn{2}{|c|}{\textbf{contributor}} \\ \hline
\textbf{URI} & \href{http://purl.org/dc/elements/1.1/contributor}{http://purl.org/dc/elements/1.1/contributor} \\ \hline
\textbf{Label} & Contributor \\ \hline
\textbf{Definition} & An entity responsible for contributing to the resource. \\ \hline

\multicolumn{2}{|c|}{\textbf{creator}} \\ \hline
\textbf{URI} & \href{http://purl.org/dc/elements/1.1/creator}{http://purl.org/dc/elements/1.1/creator} \\ \hline
\textbf{Label} & Creator \\ \hline
\textbf{Definition} & An entity primarily responsible for making the resource. \\ \hline

\multicolumn{2}{|c|}{\textbf{date}} \\ \hline
\textbf{URI} & \href{http://purl.org/dc/elements/1.1/date}{http://purl.org/dc/elements/1.1/date} \\ \hline
\textbf{Label} & Date \\ \hline
\textbf{Definition} & A point or period of time associated with an event in the lifecycle of the resource. \\ \hline
\end{tabular}
\vspace{-0.75cm}
\end{table}

\begin{table}[h!]
\renewcommand{\arraystretch}{1.25}
\centering
\caption{Corresponding \textit{terms} of the \textit{elements} in Table \ref{tab:dcmes_elements}} 
\label{tab:dcterms_terms}
\begin{tabular}{|p{2cm}|p{10.2cm}|}
\hline
\multicolumn{2}{|c|}{\textbf{contributor}} \\ \hline
\textbf{URI} & \href{http://purl.org/dc/terms/contributor}{http://purl.org/dc/terms/contributor} \\ \hline
\textbf{Label} & Contributor \\ \hline
\textbf{Definition} & An entity responsible for making contributions to the resource. \\ \hline
\textbf{Type of Term} & Property \\ \hline
\textbf{Range Includes} & \href{http://purl.org/dc/terms/Agent}{http://purl.org/dc/terms/Agent} \\ \hline
\textbf{Subproperty of} & Contributor (\href{http://purl.org/dc/elements/1.1/contributor}{http://purl.org/dc/elements/1.1/contributor}) \\ \hline

\multicolumn{2}{|c|}{\textbf{creator}} \\ \hline
\textbf{URI} & \href{http://purl.org/dc/terms/creator}{http://purl.org/dc/terms/creator} \\ \hline
\textbf{Label} & Creator \\ \hline
\textbf{Definition} & An entity responsible for making the resource. \\ \hline
\textbf{Type of Term} & Property \\ \hline
\textbf{Range Includes} & \href{http://purl.org/dc/terms/Agent}{http://purl.org/dc/terms/Agent} \\ \hline
\textbf{Subproperty of} & Creator (\href{http://purl.org/dc/elements/1.1/creator}{http://purl.org/dc/elements/1.1/creator}) \\ \hline

\multicolumn{2}{|c|}{\textbf{date}} \\ \hline
\textbf{URI} & \href{http://purl.org/dc/terms/date}{http://purl.org/dc/terms/date} \\ \hline
\textbf{Label} & Date \\ \hline
\textbf{Definition} & A point or period of time associated with an event in the lifecycle of the resource. \\ \hline
\textbf{Type of Term} & Property \\ \hline
\textbf{Has Range} & \href{http://www.w3.org/2000/01/rdf-schema#Literal}{http://www.w3.org/2000/01/rdf-schema\#Literal} \\ \hline
\textbf{Subproperty of} & Date (\href{http://purl.org/dc/elements/1.1/date}{http://purl.org/dc/elements/1.1/date}) \\ \hline

\end{tabular}

\end{table}

While implementations can use these fifteen properties in either the older format \emph{dc:} or the newer format \emph{dcterms:}, DCMI advises on its site to use the more recent one to adhere to Semantic Web's best practices. The observatory currently follows the DCTERMS structure.

\subsection{Topic overview}
\label{sec:topic_overview}

The main challenge of our research, which falls within the data integration subfield of the information retrieval domain \cite{Manning2008}, is to automate the process of ingesting and classifying STTs according to their taxonomy into the observatory.

Data integration is the process of combining data from different sources into a unified view. It involves cleansing, transforming, and consolidating data from different databases, applications, systems, or services. Data integration aims to provide meaningful and valuable information that can be easily used for analytical, operational, or transactional purposes \cite{Doan2012}. In addition to moving data from one place to another, data integration should ensure its consistency, reliability, and quality. An interesting taxonomy of data integration functions can be found online in \cite{Tatarnikova2023}.
    
ETL (Extract, Transform, and Load) explicitly focuses on the extraction, transformation, and loading phases of the data integration process. As such, ETL can be considered a subset of the data integration landscape. In data integration, ETL tools are critical for collecting data from disparate sources, transforming it into a consistent and usable format, and loading it into a target database or data warehouse. These tools enable organizations to efficiently manage, consolidate, and analyze data from disparate sources to provide a unified view for reporting and decision-making. ETL can be particularly challenging when extracting multimedia content (e.g., plain text, street addresses, URLs, logos, images, video) \cite{Mallek2024}.

\subsubsection{Smart ETL}
\label{sec:Smart_ETL_definition}
    
Our research aims to extract data from unstructured STT catalogs in PDF format, organize them in a human-readable and orderly manner, classify the STTs, and upload them to the observatory. To emphasize the use of AI to automate the STT classification process, we propose the term \textit{Smart ETL}. Our approach to automating STT classification differs from traditional machine learning (ML) classification methods in two main ways:

\begin{enumerate}[i.]
    \item The absence of a categorized dataset, typically with thousands of records, to be divided into training and test sets.
    
    \item The dataset does not consist of multiple features with categorical, ordinal, or numeric values. The classification of STTs is based solely on their textual descriptions. In addition, the model must be able to recognize implicit concepts. For example, if an STT is described as having sensors to measure crowding, this implies that it can monitor the flow of people and that the collected data can be used to adjust a tourist service. However, this information is not explicitly stated in the description; it should be inferred.
\end{enumerate}

\subsubsection{Generative AI and large language models}
\label{sec:GenAI}

Generative AI (GenAI) is a currently popular subset of AI that involves algorithms that generate new content based on their training data, including images, text, and audio. Large Language Models (LLMs) are a specific category of generative models explicitly designed to understand, generate, and manipulate human language. Among these, Transformer-based models \cite{Vaswani2017} have gained prominence, which, according to NVIDIA, \href{https://blogs.nvidia.com/blog/what-is-a-transformer-model/}{\textit{"70 percent of arXiv papers on AI posted in the last two years mention transformers" (March 25, 2022)}}. These models effectively capture context and dependencies using self-attention mechanisms and excel at NLP (Natural Language Processing) tasks, text generation, and context understanding. The transformer architecture can be divided into\footnote{Although transformer-based models are applicable beyond LLMs, the explanation of this architecture in this paper will focus exclusively on its use in LLMs.}:

\begin{itemize}
    \item \textbf{Encoder:} takes a sequence of words (like a sentence or paragraph) and processes it to understand the meaning and context of each word in relation to the others. It generates vectors of numerical representations (embeddings) that represent the input in a way the machine can understand. Among the foundational models\footnote{A foundational model refers to a large, pre-trained model that serves as a starting point or base for various specialized tasks and applications. These models are typically trained on large amounts of data and are designed to capture general patterns and features that can be fine-tuned for specific use cases.} in the encoder-only architecture, BERT (Bidirectional Encoder Representations from Transformers) \cite{Devlin2019} stands out. Learning from left and right contexts during pre-training improves its understanding of natural language. It is well suited for text classification, question answering, and other comprehension-based applications.
    
    \item \textbf{Decoder:} takes the encoded information (the embeddings) and generates a new sequence of words. It produces the output one word at a time, considering the context provided by the encoded information. GPT (Generative Pre-trained Transformer) \cite{Radford2018}, a well-known foundational model of the decoder-only architecture, generates text by predicting the next word in a sequence, making it suitable for various generative tasks such as text generation, language modeling, and conversational agents.
\end{itemize}

Combining both components results in encoder-decoder models that use both an encoder to process the input and a decoder to generate the output. The transformer model was the one that introduced this architecture. Besides that, there are other famous encoder-decoder models like T5 (Text-to-Text Transfer Transformer) \cite{Raffel2020}, which treats every NLP problem as a text-to-text task, and BART (Bidirectional and Auto-Regressive Transformers) \cite{Lewis2019}, which combines the bidirectional context of BERT with the autoregressive nature of GPT, making it versatile for both understanding and generation tasks.

Encoder-only and decoder-only models can achieve greater efficiency and effectiveness within their respective domains by specializing in specific task types. In contrast, encoder-decoder models offer versatility for tasks involving input processing and output generation.

There are two main ways of teaching or adapting an LLM to specific types of tasks:

\begin{itemize}
    \item \textbf{Fine-tuning:} involves taking a pre-trained model and further training it on a smaller, task-specific dataset. This process adjusts the model's weights based on the new data, allowing it to perform well on the specific task.
    
    \item \textbf{In-Context Learning (ICL):} involves giving the model the task explanation alongside possible examples of the expected results during inference\footnote{Inference in this context refers to the process of generating a response or prediction based on a given input.}, without additional training, allowing it to learn from these examples to complete the assignment. The model may receive zero examples (zero-shot) or a small number (few-shot) within the prompt. This approach is prominently used in decoder-only models.
\end{itemize}

Fine-tuning is ideal for achieving high performance on specific tasks when there is a sufficient amount of labeled data. However, it requires additional training, which can be resource-intensive, especially for larger models with many parameters and layers. In contrast, ICL is flexible and can quickly adapt to new tasks without additional training. This approach is critical when labeled data is scarce or computational resources for fine-tuning are insufficient.

In short, LLMs have brought significant advances to the field of NLP, from text understanding and generation to translation and summarization. For this reason, an LLM should be the solution for our STT classification approach since it should rely only on the descriptions of STTs.

        
\section{Smart extraction phase}
\label{sec:smart_extract}

In this section, we present our current extraction phase, where textual and graphical data, if available, are extracted from the catalogs.

\subsection{Extraction of elements from catalogs}

Before explaining the scraping of the STT catalogs, Table~\ref{tab:elements_extracted} outlines the specific elements extracted from each catalog. It is important to note that not all elements were extracted from every catalog due to variations in the available information.

\begin{table}[!h]
    \caption{Identification of the elements extracted from each catalog}
    \label{tab:elements_extracted}
    \fontsize{8pt}{10pt}\selectfont
    \renewcommand{\arraystretch}{1.1}
    \centering
    \begin{tabular}{|>{\centering\arraybackslash}m{5.5cm}|>{\centering\arraybackslash}m{2.0cm}|>{\centering\arraybackslash}m{1.75cm}|>{\centering\arraybackslash}m{1.75cm}|}
        \hline
        \textbf{Element} & \textbf{Catalogs of SEGITTUR} & \textbf{Catalogs of ADESTIC} & \textbf{Catalogs of EU}\\
        \hline
        STT's Name & X & X &  \\
        \hline
        STT's Description & X & X &  \\
        \hline
        Producer's Name & X & X &  \\
        \hline
        Producer's Description & X & X &  \\
        \hline
        Producer's Logo & X & X &  \\
        \hline
        Producer's URL & X & X &  \\
        \hline
        Scope of SD application & X & &  \\
        \hline
        Type of Destination & \textbf{\textcolor{resetting_color}{AI}} & X &  \\
        \hline
        Type of Solution & \textbf{\textcolor{resetting_color}{AI}} & \textbf{\textcolor{resetting_color}{AI}} &  \\
        \hline
        Producer's Phone number and Email address & & X &  \\
        \hline
        Producer's Address & & X &  \\
        \hline
        QR Code & & X &  \\
        \hline
        STT-based initiative title & & & X \\
        \hline
        STT-based initiative location & & & X \\
        \hline
        STT-based initiative description & & & X \\
        \hline
        STT-based initiative URL & & & X \\
        \hline
        STT-based initiative image(s) & & & X \\
        \hline
    \end{tabular}
\end{table}

Cells with ‘\textbf{\textcolor{resetting_color}{AI}}’ instead of ‘X’ indicate that the meaning of the extracted element was identified using AI models. Although this research mainly focuses on using LLMs to classify STTs, AI has also been used to extract content from catalogs.

The \textit{Producer Logo} and \textit{STT-based initiative image(s)} elements represent images, while the \textit{Type of Destination} in SEGITTUR catalogs and the \textit{Type of Solution} represent icons with corresponding labels. The remaining elements consist of textual data.

Regardless of the element's data type, the \href{https://pymupdf.readthedocs.io/en/latest/}{PyMuPDF} library was used to extract them. 

In the ADESTIC catalogs, explicit addresses were not available. Instead, when a landline number was provided, we attempted to identify its associated region. We extracted all Spanish prefix numbers corresponding to different Spanish regions for that purpose. Once a match was found, we used the \href{https://geopy.readthedocs.io/en/stable/}{GeoPy} Python library to obtain the corresponding coordinates to position STTs into the observatory's geographic map. Consequently, the \textit{Address} element does not represent the precise producer address but rather indicates the region from which the producer originates. The \href{https://pypi.org/project/pyzbar/}{Pyzbar} library was used for the QR code extraction.

\subsection{Leveraging AI for graphical element extraction}

Besides extracting graphical elements, obtaining their labels from the initial glossary of each catalog is crucial. This is not necessary for the \textit{Producer Logo} element since it does not represent a category. Given that the corresponding label for each icon is exclusively shown in the glossary and not on the STT pages, we utilized the \href{https://huggingface.co/microsoft/resnet-18}{resnet-18} ML model for image classification. This model is accessible via the  \href{https://www.fast.ai/}{fastai} Python library, and here is how we used it:

\begin{enumerate}[i.]
    \item The icons from the graphic glossary were extracted and saved in different image formats and resolutions.
    
    \item Data augmentation techniques were applied to increase our training data. It resulted in a dataset three times larger than the original.
    
    \item \href{https://huggingface.co/microsoft/resnet-18}{resnet-18} was fine-tuned with the augmented training data.
    
    \item Our objective is to classify glossary images (which served as our training dataset). Specifically, we aim to classify images the model has already found rather than dealing with unseen images.
\end{enumerate}

We followed this procedure for every catalog containing an iconographic glossary.

\subsection{Extracted STTs overview}

\begin{figure}[!h]
    \centering
    \includegraphics[width=1\textwidth]{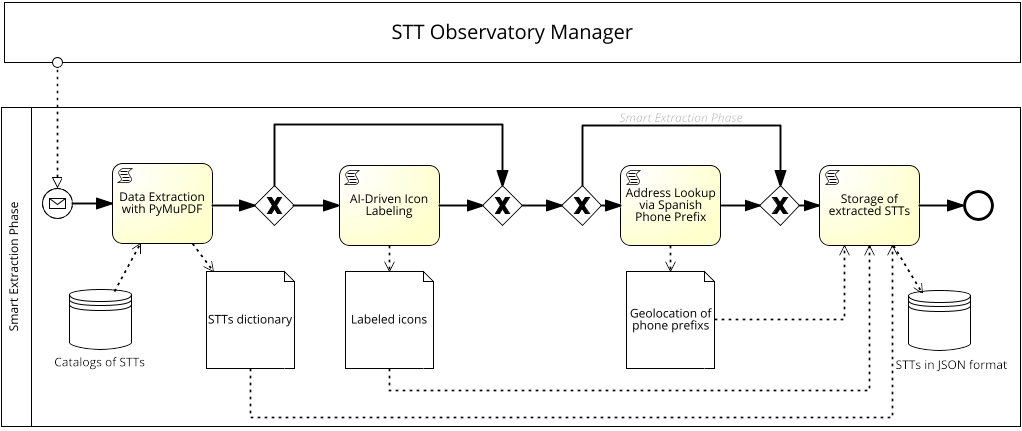}
    \caption{BPMN process model for the Smart Extraction Phase}
    \label{fig:bpmn_smart_extraction}
    \vspace{-0.5cm}
\end{figure}

Figure~\ref{fig:bpmn_smart_extraction} presents the BPMN process model for our smart extraction phase.

The code used in this section is available on the \href{https://github.com/resetting-eu/European_STT_Observatory/tree/main/Catalogs%20Extraction}{RESETTING page on GitHub}.


\section{Smart transformation phase}
\label{sec:smart_transformation}

Tables~\ref{tab:dcmes_properties} and~\ref{tab:dcmes_properties_eu} present some extracted elements that were incorporated into DCTERMS properties. Also, a DCTERMS property may comprise more than one element.


\begin{table}[ht]
    \centering
    \caption{DCTERMS property associations from Spanish catalogs}
    \label{tab:dcmes_properties}
    \fontsize{8pt}{10pt}\selectfont
    \renewcommand{\arraystretch}{1.1}
    \begin{tabular}{|l|l|}
        \hline
        \textbf{DCTERMS property} & \textbf{Element(s) Extracted} \\
        \hline
        Title &  STT Name \\
        \hline
        Description & STT Description \\
        \hline
        Creator & Producer Name; Producer Description; Producer Logo; Phone Number \\
        \hline
        Local URL & Producer URL \\
        \hline
    \end{tabular}
\end{table}

\begin{table}[ht]
    \centering
    \fontsize{8pt}{10pt}\selectfont
    \renewcommand{\arraystretch}{1.1}
    \caption{DCTERMS property associations from EU catalogs}
    \label{tab:dcmes_properties_eu}
    \begin{tabular}{|l|l|}
        \hline
        \textbf{DCTERMS property} & \textbf{Element(s) Extracted} \\
        \hline
        Title &  STT-based initiative title \\
        \hline
        Description & STT-based initiative description \\
        \hline
        Address & STT-based initiative location \\
        \hline
        Local URL & STT-based initiative URL \\
        \hline
    \end{tabular}
    
\end{table}

Omeka allows for the incorporation of the content of each item into HTML format. So, after extracting the elements, we transform them into HTML format, which provides more flexibility in how the content is laid out. In addition, since ADESTIC's catalogs are in Spanish, the \href{https://pypi.org/project/deep-translator/}{deep-translator} library was used to translate them into English.

In the Transformation phase, LLMs were used for removing duplicates (Section~\ref{sec:duplicates_removal}) and classifying STTs (Section~\ref{sec:stts_classification}). Before diving into the tasks, we will justify the choice of the LLMs.

\subsection{Choice of LLMs}
\label{sec:choice_of_llms}

Due to resource constraints, fine-tuning an LLM with satisfactory performance was not feasible. Hosting an LLM solely for inference was also not a viable option. Consequently, model selection should exclusively consider decoder-only models since they are the only type accessible online without needing local execution.

\href{https://huggingface.co/}{Hugging Face} is a French-American company and an open-source community focusing on NLP models and tools. They are renowned for their \href{https://huggingface.co/docs/transformers/index}{Transformers library}, an open-source platform that offers user-friendly interfaces to cutting-edge pre-trained NLP models. Besides that, they have also developed \href{https://huggingface.co/chat/}{HuggingChat}, e.g., a free AI-powered conversational agent with the latest NLP models. \href{https://github.com/Soulter/hugging-chat-api}{Hugchat} is an unofficial Python API for HuggingChat, but notably, Hugging Face's Chief Technology Officer (CTO) has expressed appreciation for the project on their GitHub page.

For the duplicates removal task we used, through Hugchat, the \href{https://huggingface.co/mistralai/Mixtral-8x7B-Instruct-v0.1}{\textit{mistralai/Mixtral-8x7B-Instruct-v0.1}}. At the time of execution, this model was Hugchat's default LLM and one of the best LLMs available.

Regarding the primary subject of this paper, the LLM-based STT classification, we employed different LLMs to assess their performance: 
\textit{\href{https://copilot.microsoft.com/}{Microsoft Copilot}}, \textit{\href{https://huggingface.co/NousResearch/Nous-Hermes-2-Mixtral-8x7B-DPO}{NousResearch/Nous-Hermes-2-Mixtral-8x7B-DPO}}, \textit{\href{https://huggingface.co/mistralai/Mixtral-8x7B-Instruct-v0.1}{mistralai/Mixtral-8x7B-Instruct-v0.1}}, and \textit{\href{https://huggingface.co/meta-llama/Meta-Llama-3.1-70B-Instruct}{meta-llama/Meta-Llama-3.1-70B-Instruct}} (the last three from Hugchat).

\subsection{Duplicates removal}
\label{sec:duplicates_removal}
Some redundancies occurred since both ADESTIC and SEGITTUR catalogs focus on Spanish STTs. However, when ADESTIC is translated into English, the STT names may not match those in the SEGITTUR catalogs. Furthermore, solution types in different catalogs may not share identical names, even if they represent the same concept, and a type in one catalog may cover several types in another. As a result, manual associations have been established between solution types. However, some solution types remain unassociated because there is no possible association with the types in another catalog. The manual association is available  \href{https://github.com/resetting-eu/European_STT_Observatory/blob/main/Catalogs%20Extraction/Solutions%20Types%20(SEGITTUR%20and%20ADESTIC)/Type_of_Solution_Association%20(EN).json}{here}.

Therefore, the duplicate removal process works as follows:
\begin{enumerate}[i.]
    \item The STT and producer names between two STTs are compared.
    
    \item If no match is found, the STTs associated with similar solution types are retrieved. For instance, when searching in the ADESTIC V2 catalog for duplicates of a STT from the SEGITTUR 2022 catalog with the solution type \textit{"Efficient Management: Energy"}, we will be looking for STTs having the solution type \textit{"Efficient Management: water, air, energy or waste"} (Box \ref{box:similar}).

\vspace{0.1cm}
\noindent\begin{minipage}{\textwidth}
\captionof{colorbox}{Example of similar solution types}
\label{box:similar}
\end{minipage}
\vspace{-0.4cm}
\begin{tcolorbox}[colframe=lightgray, colback=blue!10, breakable]
\fontsize{8pt}{11pt}\selectfont
\fontfamily{phv}\selectfont
\begin{tabbing}
\hspace{0.30cm} \= \hspace{1cm} \= \hspace{2cm} \= \kill
\{ \\
   \> "Segittur 2022": [ 
   \end{tabbing}
   \begin{tabbing}
   \hspace{0.80cm} \= \hspace{1.5cm} \= \kill
      \> "Efficient Management: Energy", \\
      \> "Efficient Management: Water", \\
      \> "Efficient Management: Air quality", \\
      \> "Efficient Management: Waste" 
    \end{tabbing}
    \begin{tabbing}
    \hspace{0.30cm} \= \hspace{1cm} \= \hspace{2cm} \= \kill
    \> ], \\
    \> "Segittur 2023": "Efficient resource management: 
                    water/energy/waste/air quality", \\
   \> "Adestic V1": "Efficient Management: water, air, energy 
                 or waste", \\
   \> "Adestic V2": "Efficient Management: water, air, energy 
                 or waste" \\
\}
\end{tabbing}
\end{tcolorbox}

Another limitation has been prevented; we will explain it through an example for understandability's sake. When searching within the SEGITTUR 2023 catalog for duplicates of a STT from the ADESTIC V1 catalog with \textit{"Accessibility"} as the solution type, and if the SEGITTUR 2023 catalog lacks that solution type (Box \ref{box:not_available}), all solution types without an association between these two catalogs are retrieved (Box \ref{box:unassociated}). This process helps identify solutions that are the same but were categorized differently by the catalog administrators.

    In a real application, the "\{...\}" in Box \ref{box:unassociated} would be replaced by the remaining types, that have no identified association.
    
\item Subsequently, each STT will undergo comparison using the LLM. The \href{https://github.com/resetting-eu/European_STT_Observatory/blob/main/Catalogs%20Extraction/Prompt_To_Identfy_Repetitions.txt}{prompt provided in a zero-shot setup} is available on the European STT Observatory GitHub page, accompanied by an \href{https://github.com/resetting-eu/European_STT_Observatory/blob/main/Catalogs%20Extraction/Prompt_To_Identfy_Repetitions_Example.txt}{application example}.

\noindent\begin{minipage}{\textwidth}
\captionof{colorbox}{Example of solution type not available in all catalogs}
\label{box:not_available}
\end{minipage}
\vspace{-0.4cm}
\begin{tcolorbox}[colframe=lightgray, colback=blue!10, breakable, label={box:example}]
\fontsize{8pt}{11pt}\selectfont
\fontfamily{phv}\selectfont
\begin{tabbing}
\hspace{0.30cm} \= \kill
\{ \\
   \> "Segittur 2022": "Accessibility", \\
   \> "Segittur 2023": "", \\
   \> "Adestic V1": "Accessibility", \\
   \> "Adestic V2": "Accessibility" \\    
\}
\end{tabbing}
\end{tcolorbox}


\vspace{-0.2cm}
\noindent\begin{minipage}{\textwidth}
\captionof{colorbox}{Example of unassociated solution types}
\label{box:unassociated}
\end{minipage}
\vspace{-0.4cm}
\begin{tcolorbox}[colframe=lightgray, colback=blue!10, breakable]
\fontsize{8pt}{11pt}\selectfont
\fontfamily{phv}\selectfont
\begin{tabbing}
\hspace{0.30cm} \= \hspace{1cm} \= \hspace{2cm} \= \kill
\{ \\
   \> "Segittur 2022": "Accessibility", \\
   \> "Segittur 2023": [ 
   \end{tabbing}
   \begin{tabbing}
   \hspace{0.60cm} \= \hspace{1.5cm} \= \kill
     \> \{ 
    \end{tabbing}
    \begin{tabbing}
   \hspace{0.80cm} \= \hspace{2cm} \= \kill
       \> "Segittur 2022": "", \\
       \> "Segittur 2023": "Intelligent Signage/Totems/Tourism Signage", \\
       \> "Adestic V1": "", \\
       \> "Adestic V2": "" 
    \end{tabbing}
    \begin{tabbing}
   \hspace{0.60cm} \= \hspace{1cm} \= \kill
     \> \}, \{...\}], 
      \end{tabbing}
      \begin{tabbing}
\hspace{0.30cm} \= \hspace{1cm} \= \kill
   \> "Adestic V1": "Accessibility", \\
   \> "Adestic V2": "Accessibility" \\
\}
\end{tabbing}
\end{tcolorbox}

\end{enumerate}

Through the application of this process, unanticipated outcomes emerged. The LLM successfully identified identical STTs and revealed instances where an STT, originally comprising multiple functionalities within one catalog, was fragmented into several distinct STTs in another catalog. We decided to keep those that were separate and eliminate those that included several STTs.

The number of duplicates identified were:
\begin{itemize}
    \item \textbf{By STT and producer names comparison:} 156
    \item \textbf{By LLM evaluation:} 57
\end{itemize}

The code created for this task is available \href{https://github.com/resetting-eu/European_STT_Observatory/blob/main/Catalogs%20Extraction/Get_Repetitions_Among_Catalogs.ipynb}{here}.

\subsection{STTs classification}
\label{sec:stts_classification}

As noted above, the European Commission catalogs were not used as a data source for this task because they do not specifically present STTs.

While several LLMs were evaluated, only Microsoft Copilot in Precise Mode was employed to define the necessary prompts. This decision was influenced by the fact that, at that time, the GPT-4 Turbo-powered Copilot Precise mode was considered one of the top-performing models.

\subsubsection{Needle in a haystack challenge}
\label{sec:needle}

Although the context length of Copilot is known, we conducted a test to determine the model's ability to detect fine details within the provided context. This detection is essential for our task, so we can know the model's limit to understand the taxonomy and the few-shot examples provided. The test was a needle in a haystack challenge, where the haystack was the text of an extracted book and the needle was a short sentence out of the context. The needle was randomly placed in the book text, and the LLM was asked to find it.

\begin{itemize}
    \item \textbf{Haystack:} \href{https://www.gutenberg.org/ebooks/1105}{The Sonnets by William Shakespeare}
    \item \textbf{Needle:} \textit{"Portugal's national team became European champions in 2016 against France. The final was played in Paris and the final score was 1-0 after extra time."}
\end{itemize}

The prompt template provided to the model is the one in Box \ref{box:needle_in_haystack}.

\noindent\begin{minipage}{\textwidth}
\captionof{colorbox}{Prompt template for the needle in the haystack challenge}
\label{box:needle_in_haystack}
\end{minipage}
\vspace{-0.5cm}
\begin{tcolorbox}[colframe=lightgray, colback=blue!10]
\fontsize{8pt}{11pt}\selectfont
\fontfamily{phv}\selectfont
\begin{tabbing}
\hspace{0.30cm} \= \kill
\parbox{\linewidth}{
"Let's do the needle in a haystack challenge. In the following 
text, you have to find the needle, which is a sentence out of 
context. Good luck! \\ 
The text is: "\#\#\#TEXT TO REPLACE\#\#\#" \\
You have to return this JSON object. The 'Needle' value is the 
sentence out of context. If you don't find the needle the 
value must be 'NOT FOUND': \{"Needle": <SENTENCE>\}." 
}
\end{tabbing}
\end{tcolorbox}
\vspace{-0.1cm}

We initiated the test with a prompt character limit of 10,000.
If the LLM responded correctly, we increased the limit by 500 characters. However, if the model provided incorrect answers 6 times consecutively, the test was concluded. In addition, we calculated the quartile in which the needle was randomly inserted in the text of the book so that we could get a better idea of the influence on the model's response.

\begin{figure}[ht!]
    \centering
    \includegraphics[height=0.7\textheight]{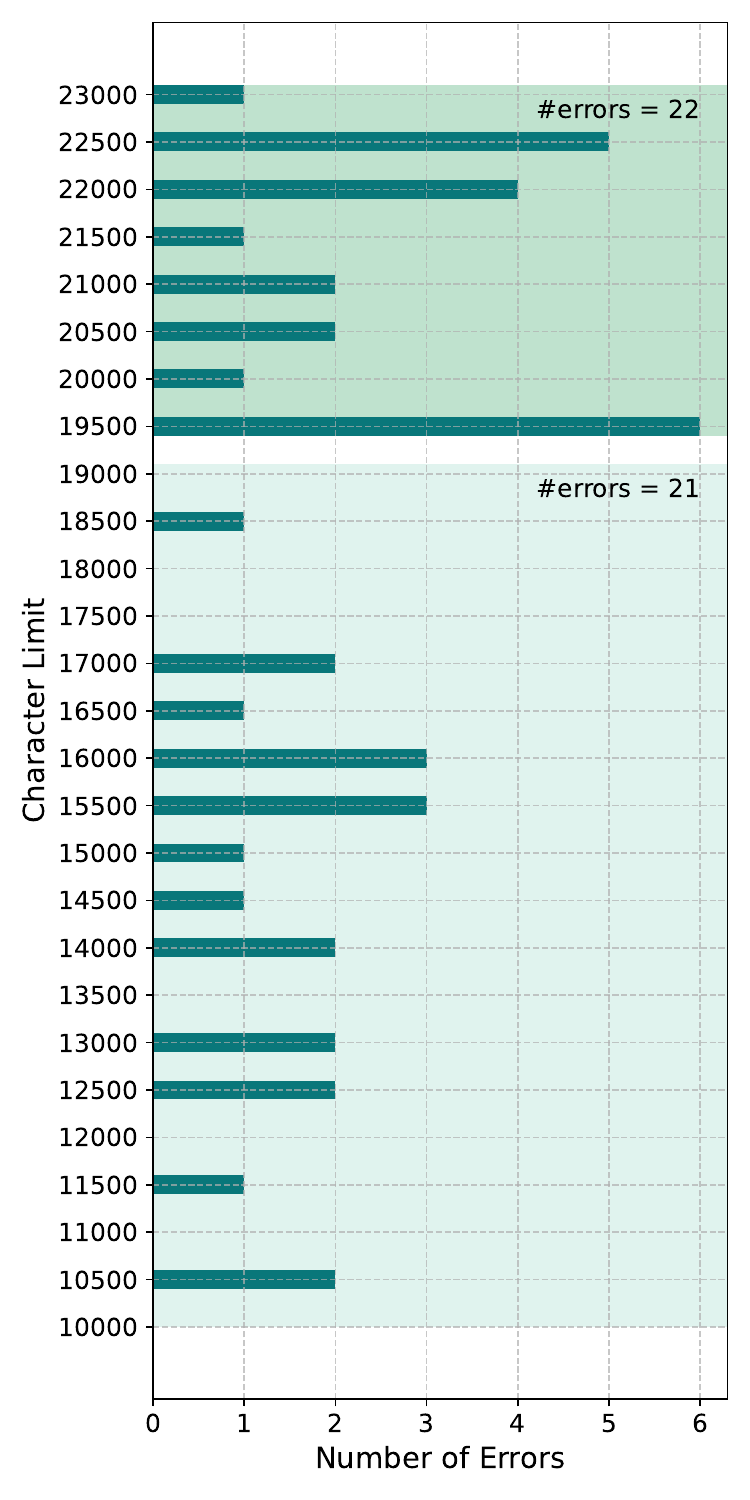}
    \caption{Character limit-based error analysis}
    \label{fig:character_limit_errors}
    \vspace{-1.15cm}
\end{figure}

We conducted a total of four tests. However, all tests were prematurely terminated due to reaching the Copilot online character limit. As a result, we were unable to accurately determine the precise character limit at which the LLM can still successfully locate the needle. Despite this limitation, we were able to derive several valuable insights:

\begin{itemize}
    \item Copilot may be capable of finding the needle in the haystack when the character limit is above 23,000.

    \item Figure~\ref{fig:quartile_error} shows that Copilot performs better at detecting small details in the second half of the prompt, as evidenced by the lower percentage of errors in the third and fourth quartiles. Note again that the needle was placed randomly, so \textit{n} is not the same for all quartiles.

\begin{figure}[ht!]
    \centering
    \includegraphics[width=.75\textwidth]{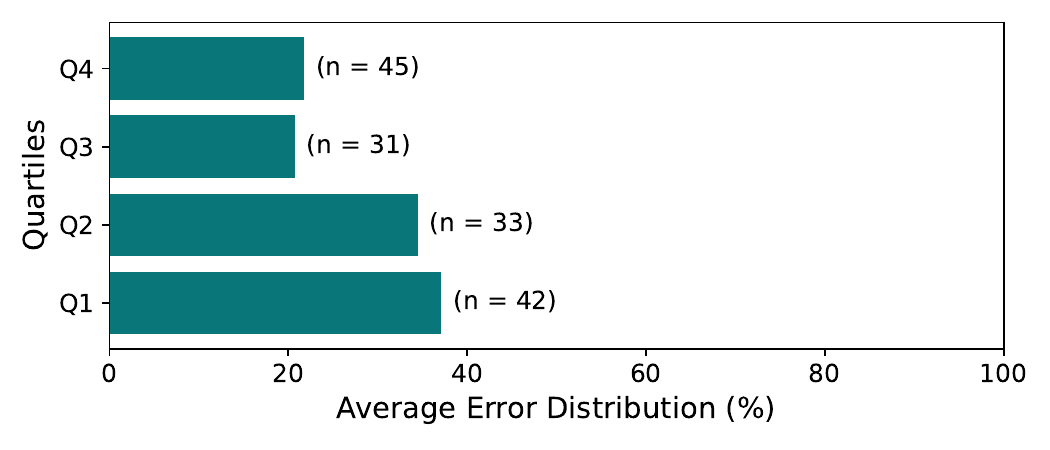}
    \caption{Quartile-based analysis of error distribution}
    \label{fig:quartile_error}
    \vspace{-0.5cm}
\end{figure}

    \item The requested JSON object was provided in 85\% of the responses, for a total of 151 responses. This higher accuracy may confirm the previous point that the model is better at detecting small details in the second part of the prompt, since the JSON request is the final statement. However, for a comprehensive analysis, the JSON request should be placed in different parts of the prompt for comparison, which was not done in this research as it was not the primary focus.
    
    \item Figure~\ref{fig:character_limit_errors} illustrates that the probability of errors increases as the maximum character limit increases. The 19,500-23,000 range had almost as many errors as the 10,000-19,000 range, with 22 errors compared to 21. It is important to note that the test was terminated after six errors, not after the first error. This approach allows us to estimate the likelihood of Copilot making errors at each character limit. 
\end{itemize}

\setlength{\belowcaptionskip}{0pt}  

\subsubsection{Classification methodology}

Based on these results, we decided on a maximum character limit of 15,000 characters for the STT classification task. For that, we employed both training and testing datasets. However, due to the absence of a large categorized dataset, we manually categorized some STTs as follows:

\begin{itemize}

    \item \textbf{STTs Selection:} We randomly selected STTs from each Spanish catalog, since only those catalogs contained STTs. We extracted an STT from one catalog and categorized it, then repeated the process for the other catalogs. 
    
    \item \textbf{Dataset Sizes:} After reviewing all four catalogs, we returned to the first one and repeated the process until we created two datasets, each containing STTs that covered all the categories in the taxonomy. In one test, one dataset serves as the training set and the other as the testing set. In the subsequent test, the roles are reversed, with the training dataset becoming the testing set and vice versa. Since an STT can cover multiple categories, the size of the datasets is smaller than the number of categories. Thus, Dataset A contains 8 STTs, and Dataset B also contains 9 STTs.
\end{itemize}

The prompts for this task were created through an iterative approach, with changes or additions made in each iteration. Three prompts were utilized: one to describe the STT taxonomy, one for the few-shots, and another to introduce each STT to be classified. We verified the need to update the prompts through the answers given by the LLM about the STTs given to be classified. The answers let us ascertain when a definition was unclear, namely when the LLM confused two or more categories. When it was necessary to shorten or clarify a sentence or paragraph in the prompts, we found it advantageous to use the LLM for this task simply by asking the LLM to do so and providing the respective text to be updated.

\begin{itemize}
    \item \textbf{Prompt of Taxonomy Definition:} the earlier versions of this proposal included an outdated taxonomy. We recognized the need for an update in the \textit{"(Part of) the tourist offer"} domain because the LLM struggled to fully understand it, and we concluded that people might face the same difficulty. A report on the iterations made while developing this prompt is available \href{https://github.com/resetting-eu/European_STT_Observatory/blob/main/Progress%20of%20the%20Prompt%20of%20STT%20Taxonomy%20Definition.pdf}{online}.
    
    \item \textbf{Few-shot Prompt:} we differentiated the few-shot examples from the STTs to be classified using initial identifiers in the prompts: \textit{"\#\#\#EXAMPLE\#\#\#"} or \textit{"\#\#\#Classification\#\#\#"}. A structure identical to the one in Box \ref{box:few_shot} was used for each shot.

    Although it does not always work (i.e., the LLM did not comply with the request), the last statement in the prompt is essential to prevent the LLM from giving long answers to the few-shot examples in most cases. In addition, for the classification justification, we created an expression for each STT category so that we could repeat it in the justification every time that category was selected, thus achieving consistency and repetition between the few-shot examples, and trying to make the job of interpretation easier for the LLM. An example is:

    \textbf{Tourist Experience:} \textit{“It is “Tourist Experience” because it requires the tourist’s active participation (i.e., …) and it's not about the planning, organization, or execution of the activity or future activities.”}

    This sentence is repeated in the justifications each time this category is selected, with "…" replaced by information that classifies the STT as "Tourist Experience". If the combination of two or more few-shots did not exceed the limit identified in the Needle in a Haystack test, they were sent in the same prompt.

     \vspace{0.2cm}
    \noindent\begin{minipage}{\textwidth}
    \captionof{colorbox}{Few-shot example structure}
    \label{box:few_shot}
    \end{minipage}
    \vspace{-0.5cm}
    \begin{tcolorbox}[colframe=lightgray, colback=blue!10, breakable]
    \fontsize{8pt}{11pt}\selectfont
    \fontfamily{phv}\selectfont
    \begin{tabbing}
    \hspace{0.30cm} \= \kill
    \#\#\#EXAMPLE\#\#\# \\
    \{ \\
        \> "Solution Name": "...", \\
        \> "Description": "..." \\    
    \}, \\
    \{ \\
       \> "Classification": [...], \\
       \> "Justification": "..." \\
    \} \\
    This message is only used for context and does not require 
    a response.
    \end{tabbing}
    \end{tcolorbox}

    \item \textbf{Prompt to introduce the STT to be classified:} our initial idea was to use a prompting approach where the LLM would assume the roles of three different appropriate experts to solve the classification problem. At each step of reasoning, the experts would share their thoughts with the group, and each expert would score their peers’ responses on a scale from 1 (highly unlikely) to 5 (highly likely). Finally, the LLM would analyze the three expert analyses and provide either a consensus definition or its best guess solution. However, this approach proved too complex for the LLM to complete the assigned task, so we adopted a different strategy. Instead of asking the LLM to identify and behave as three different experts suitable for the problem, we instructed the model to assume the roles of three experts, each specializing in a different application domain of the taxonomy. Each expert assigned a score from 1 (highly unlikely) to 5 (highly likely) and an explanation to their respective categories and subcategories, if applicable. The LLM then identified the STT categories with scores of 4 or higher, or those with the highest scores if none reached 4. At the end, the LLM should return a JSON object with a key for each domain and the corresponding identified categories as values (Box \ref{box:stt_classification_json}). 

    \vspace{0.2cm}
    \noindent\begin{minipage}{\textwidth}
    \captionof{colorbox}{STT classification JSON}
    \label{box:stt_classification_json}
    \end{minipage}
    \vspace{-0.5cm}
    \begin{tcolorbox}[colframe=lightgray, colback=blue!10, breakable]
    \fontsize{8pt}{11pt}\selectfont
    \fontfamily{phv}\selectfont
    \{
        \begin{tabbing}
        \hspace{0.70cm} \= \kill
        \> "(Part of) the touristic offer": [...], \\
        \> "Marketing": [...], \\
        \> "Management \& Operations": [...]
        \end{tabbing}
    \}
    \end{tcolorbox}
    
    This approach was more viable and performed better. In addition to the prompt, the STT to be classified was provided in a JSON object with "Solution Name" and "Description" as keys and their respective texts as values. A report of the prompt iterations is available \href{https://github.com/resetting-eu/European_STT_Observatory/blob/main/Progress%20of%20the%20Prompt%20to%20introduce%20the%20STT%20to%20be%20classified.pdf}{online}.
  
\end{itemize}

A maximum of ten prompts were sent per context length, i.e., the same conversation with an LLM. These prompts were divided into the \textit{taxonomy definition prompt}, the \textit{few-shot prompts}, and the \textit{STTs to be classified prompts}. The number of prompts sent in the same conversation varied depending on the possible combinations of few-shots within a single prompt (see explanation of \textbf{few-shot prompt}) and the split of the test dataset. The test dataset was split into two parts since the total number of STTs to be classified, combined with the other prompts, exceeded ten prompts. Additionally, dataset B has an odd number of entries, resulting in an additional prompt being sent in one of the conversations.

As mentioned, the datasets were manually labeled. However, since this taxonomy is new, our experience in classifying STTs according to it was not fully mature. Therefore, during the prompt refinement process, we reviewed the justification provided when the LLM suggested a category for an STT that we had not initially assigned. If it made sense, we added that category to the STT. The opposite also happened when we had a category incorrectly assigned.

The results obtained, the limitations, and the negative or unexpected findings of the STT classification are presented in Section~\ref{sec:results}.

\section{Load phase}
\label{sec:load_phase}

\begin{figure}[h!]
    \centering
    \includegraphics[width=1.0\textwidth]{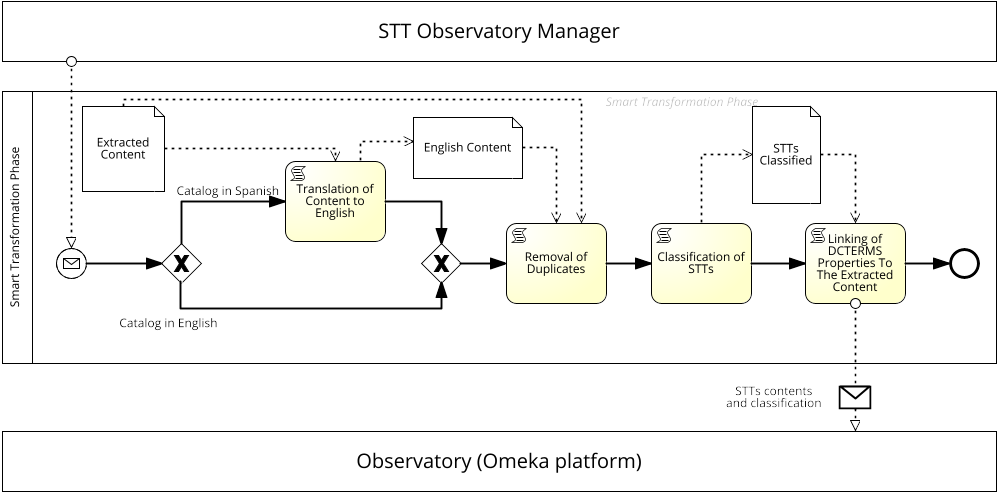}
    \caption{BPMN process model for the Smart Transformation and Load Phases}
    \label{fig:bpmn_smart_transform}
\end{figure}

This was the only phase where AI was not used. One of the reasons for choosing Omeka.net is that it includes an API that allows us to insert STTs into the observatory automatically. It works via HTTP requests with the data field in JSON format, and for tasks related to inserting, updating, and deleting content, it requires an API key that only the observatory managers have. We created a Python class called OmekaAPI to facilitate the API processes. The class does not contain all the features that Omeka.net allows, but only the ones we need, and is available \href{https://github.com/resetting-eu/European_STT_Observatory/blob/main/Omeka/OmekaAPI.py}{here}.
To help sum up these last two sections, Figure~\ref{fig:bpmn_smart_transform} presents the BPMN process of the Smart Transformation Phase and the Load Phase applied in this research.

\section{Results and discussion}
\label{sec:results}

Here, we will focus on presenting the results of the principal task of our Smart ETL process, i.e., STT classification. Each test corresponds to the classification of an STT, following a black-box approach due to the LLMs' architecture, and its response is considered valid as long as the returned JSON contains the required keys and the STT categories identified as values. We now discuss the results of the STTs classification for each LLM separately:

\begin{itemize}
    \item \textit{\textbf{\href{https://huggingface.co/NousResearch/Nous-Hermes-2-Mixtral-8x7B-DPO}{NousResearch/Nous-Hermes-2-Mixtral-8x7B-DPO}:}} The LLM failed the tests due to its difficulty in understanding the taxonomy and the classification task. This was evident after completing the first full set of tests with dataset A, as the results did not match the format of the expected results and showed a poor understanding of the taxonomy.
    
    \item \textit{\textbf{\href{https://huggingface.co/mistralai/Mixtral-8x7B-Instruct-v0.1}{mistralai/Mixtral-8x7B-Instruct-v0.1}:}} We discarded the use of this LLM due to its recurrent hallucination, i.e., most of the classified STTs were assigned to non-taxonomy categories such as \textit{"Tourist Information Systems"}, \textit{"Centralized Database \& Distribution System"}, \textit{"Booking Systems"}, or \textit{"Appointment Manager"}, among others. Running two full sets of tests, one with dataset A and the other with dataset B, was sufficient to conclude that the reasoning abilities of this model were inadequate for this task.
    
    \item \textit{\textbf{\href{https://huggingface.co/meta-llama/Meta-Llama-3.1-70B-Instruct}{meta-llama/Meta-Llama-3.1-70B-Instruct}:}} This LLM ``understood'' the required task. However, the output for the \textit{(Part of) the Touristic Offer} domain was sometimes inaccurate. Like in the previous model, but less recurrently, it hallucinated, categorizing an STT with non-taxonomy categories like \textit{"Tourist Infrastructure"} or \textit{"Parking Infrastructure"}. At other times, it only returned the first-level category, such as \textit{"Tourist Experience"} or \textit{"Tourist Experience Lifecycle Management"}. Another minor issue was splitting categories like \textit{"Tourist Experience Lifecycle Management, Building Block"} into separate categories. Despite this minor error, we considered the responses to be valid for the metrics, as it is possible to infer implicitly the category assigned. Table~\ref{tab:metrics_llama70b} shows the average precision, recall and F1 scores for each domain, together with the overall F1 score. Each dataset was tested twice on different days and times to assess performance variability.

\begin{table}[ht]
    \centering
    \renewcommand{\arraystretch}{1.2}
    \caption{Performance Metrics of \textit{meta-llama/Meta-Llama-3.1-70B-Instruct} Model}
    \label{tab:metrics_llama70b}
    \begin{tabular}{|c|c|c|c|c|c|c|}
    \hline
    \textbf{\#} & \textbf{Testing Dataset} & \textbf{Total F1} & \textbf{Domain} & \textbf{Precision} & \textbf{Recall} & \textbf{F1} \\
    \hline
    \multirow{3}{*}{1} & \multirow{3}{*}{A} & \multirow{3}{*}{\parbox{0.09\textwidth}{\centering 0.50}} & (Part of) the Touristic Offer & 0.13 & 0.19 & 0.15 \\   
    &  &  & Marketing & 0.88 & 0.77 & 0.81 \\
    &  &  & Management \& Operations & 0.75 & 0.44 & 0.54 \\ \hline

    \multirow{3}{*}{2} & \multirow{3}{*}{A} & \multirow{3}{*}{\parbox{0.09\textwidth}{\centering 0.61}} & (Part of) the Touristic Offer & 0.19 & 0.31 & 0.23 \\ 
    &  &  & Marketing & 1.0 & 0.85 & 0.90 \\ 
    &  &  & Management \& Operations & 0.81 & 0.69 & 0.71 \\ \hline

    \multirow{3}{*}{1} & \multirow{3}{*}{B} & \multirow{3}{*}{\parbox{0.09\textwidth}{\centering 0.47}} & (Part of) the Touristic Offer & 0.17 & 0.22 & 0.19 \\    
    &  &  & Marketing & 0.67 & 0.54 & 0.57 \\
    &  &  & Management \& Operations & 0.74 & 0.63 & 0.64 \\ \hline

    \multirow{3}{*}{2} & \multirow{3}{*}{B} & \multirow{3}{*}{\parbox{0.09\textwidth}{\centering 0.37}} & (Part of) the Touristic Offer & 0.0 & 0.0 & 0.0 \\
    &  &  & Marketing & 0.61 & 0.54 & 0.54 \\
    &  &  & Management \& Operations & 0.70 & 0.57 & 0.59 \\ \hline
    \end{tabular}
\end{table}

    \item \textit{\textbf{\href{https://copilot.microsoft.com/}{Microsoft Copilot}:}} This model achieved the best results but also had the most hallucinations (being the more tested model may be a consequence). The \textit{(Part of) the Touristic Offer} domain continued to present difficulties. Common errors included: 
    \begin{itemize}
        \item Treat the first-level categories (\textit{"Tourist Experience"} or \textit{"Tourist Experience Lifecycle Management"}) as domains. In the returned JSON, these categories were incorrectly placed as keys with the second-level categories as their values. Despite this, these answers were accepted because they were easily associated with the corresponding domain.
        
        \item In some tests, there was repetition in the answers. Specifically, the LLM assigned the same categorization to multiple STTs. As a result, these tests were considered invalid and excluded from the metric calculations. Since the LLM is available online, we chose not to include these invalid tests in the metrics, as they reflect instances where the LLM hallucinated or repeatedly gave the same answer, potentially indicating times when Copilot was more overloaded and performing sub-optimally.
        
        \item Occasionally, the model used information from previously classified STTs within the same context window to justify the classification of the current STT.
        
        \item Another error, which occurred only once but highlighted the hallucination issues, involved a test where instead of returning the requested JSON, the LLM returned a JSON with the categories as keys and the STT name as a value for each key.
    \end{itemize}
    
    Table~\ref{tab:metrics_copilot} shows the average precision, recall and F1 scores for each domain, together with the combined F1 score for valid tests. Each dataset was tested twice at different times/days to assess performance variability.

 \begin{table}[ht!]
    \centering
    \renewcommand{\arraystretch}{1.2}
    \caption{Performance Metrics of the \textit{Microsoft Copilot} Model}
    \label{tab:metrics_copilot}
    \begin{tabular}
    {|c|c|>{\centering\arraybackslash}p{0.09\textwidth}|c|>{\centering\arraybackslash}p{0.10\textwidth}|>{\centering\arraybackslash}p{0.08\textwidth}|>{\centering\arraybackslash}p{0.05\textwidth}|}
    \hline
    \textbf{\#} & \textbf{Testing Dataset} & \textbf{Total F1} & \textbf{Domain} & \textbf{Precision} & \textbf{Recall} & \textbf{F1} \\
    \hline
    \multirow{3}{*}{1} & \multirow{3}{*}{A} & \multirow{3}{*}{\parbox{0.09\textwidth}{\centering 0.76}} & (Part of) the Touristic Offer & 0.88 & 0.81 & 0.83 \\    
    &  &  & Marketing & 0.94 & 0.92 & 0.90 \\ 
    &  &  & Management \& Operations & 0.52 & 0.63 & 0.56 \\ \hline

    \multirow{3}{*}{2} & \multirow{3}{*}{A} & \multirow{3}{*}{\parbox{0.09\textwidth}{\centering 0.75}} & (Part of) the Touristic Offer & 0.75 & 0.69 & 0.71 \\    
    &  &  & Marketing & 0.81 & 0.83 & 0.81 \\ 
    &  &  & Management \& Operations & 0.65 & 0.88 & 0.73 \\ \hline

    \multirow{3}{*}{1} & \multirow{3}{*}{B} & \multirow{3}{*}{\parbox{0.09\textwidth}{\centering 0.53}} & (Part of) the Touristic Offer & 0.33 & 0.22 & 0.26 \\    
    &  &  & Marketing & 0.67 & 0.78 & 0.70 \\ 
    &  &  & Management \& Operations & 0.56 & 0.85 & 0.63 \\ \hline

    \multirow{3}{*}{2} & \multirow{3}{*}{B} & \multirow{3}{*}{\parbox{0.09\textwidth}{\centering 0.68}} & (Part of) the Touristic Offer & 0.78 & 0.67 & 0.70 \\    
    &  &  & Marketing & 0.67 & 0.57 & 0.61 \\ 
    &  &  & Management \& Operations & 0.76 & 0.76 & 0.76 \\ \hline
    \end{tabular}
    \vspace{-0.5cm}
\end{table}
   
\end{itemize}



One of the limitations of this research is that the prompts were only refined using Microsoft Copilot and were not customized for each model used. As a result, the prompts may have been over-optimized for Microsoft Copilot, potentially leading to suboptimal performance in other models. In addition, because the models were used online rather than locally, we were affected by the variability of other users' interactions with the model. This left us uncertain about the model's capabilities during testing, making it unclear whether occasional negative results were due to potential model overload or our methodology.

\section{Verification and validation}
\label{sec:verification_and_validation}

\subsection{Demonstration}
\label{sec:demonstration}

For \textbf{online demonstration}, we created a \href{https://sites.google.com/iscte-iul.pt/resetting-project/home/smart-tourism-observatory?authuser=0}{web site} containing all the information related to the observatory. This site includes a \href{https://sites.google.com/iscte-iul.pt/resetting-project/home/smart-tourism-observatory/UserManual?authuser=0}{user manual} detailing the Observatory's functionalities, a \href{https://youtu.be/bMTk8xYevhk}{video demo}, and \href{https://sites.google.com/iscte-iul.pt/resetting-project/home/smart-tourism-observatory/TechnicalDocumentation?authuser=0}{technical documentation} summarizing the main technical aspects of the project, such as Smart ETL, Large Language Models, the methodology used, as well as the tools and a link to the \href{https://github.com/resetting-eu/European_STT_Observatory}{GitHub repository} with the implemented code.

Several \textbf{onsite demonstrations} also took place:

\begin{itemize}
    \item At the \href{https://sites.google.com/iscte-iul.pt/resetting-project/home/dissemination/IscteKnowledgeInnovation}{inauguration of the new ``Iscte - Knowledge and Innovation'' building} in Lisbon on November 20, 2023. This demonstration was attended by the then Prime Minister, the Minister for Science, Technology and Higher Education, the Minister for Cohesion, the Minister for Culture, two Secretaries of State, a large entourage representing the most diverse sectors, and the media.

    \item At the \href{https://sites.google.com/iscte-iul.pt/resetting-project/home/events/hackathon}{RESETTING Hackathon}, held at Tecnoparc in Reus, Spain, on May 2024, which focused on "TourismTech solutions to address the challenges of European tourism". During this event, each tool developed in the RESETTING project, including the Observatory, was presented, accompanied by a poster. All posters are available \href{https://sites.google.com/iscte-iul.pt/resetting-project/home/events/2nd_posters_demos_workshop?authuser=0}{here}.

    \item At the \href{https://sites.google.com/iscte-iul.pt/resetting-project/home/events/final_conference/full-program?authuser=0}{RESETTING Final Conference}, held at Auditori Diputació in Tarragona, Spain, also on May 2024. During this event, the tools and posters were presented to the conference attendees, including some SME owners, our desired main target audience, who received funding from the RESETTING project.

\end{itemize}

\subsection{Evaluation}
\label{sec:evaluation}

\subsubsection{Continuity}

We sent a survey to different smart tourism stakeholders to draw conclusions on the usability and usefulness of the European STTs Observatory and get insights on possible missing features. The survey asked about the types of professionals for whom the Observatory could be helpful, how they rated the demo, user manual, and technical documentation, how likely they would be to use the Observatory in the future, and what additional features they would like to see implemented.

\begin{figure}[ht!]
    \centering
    \begin{minipage}[b]{0.65\textwidth}
        \centering
        \includegraphics[width=\textwidth]{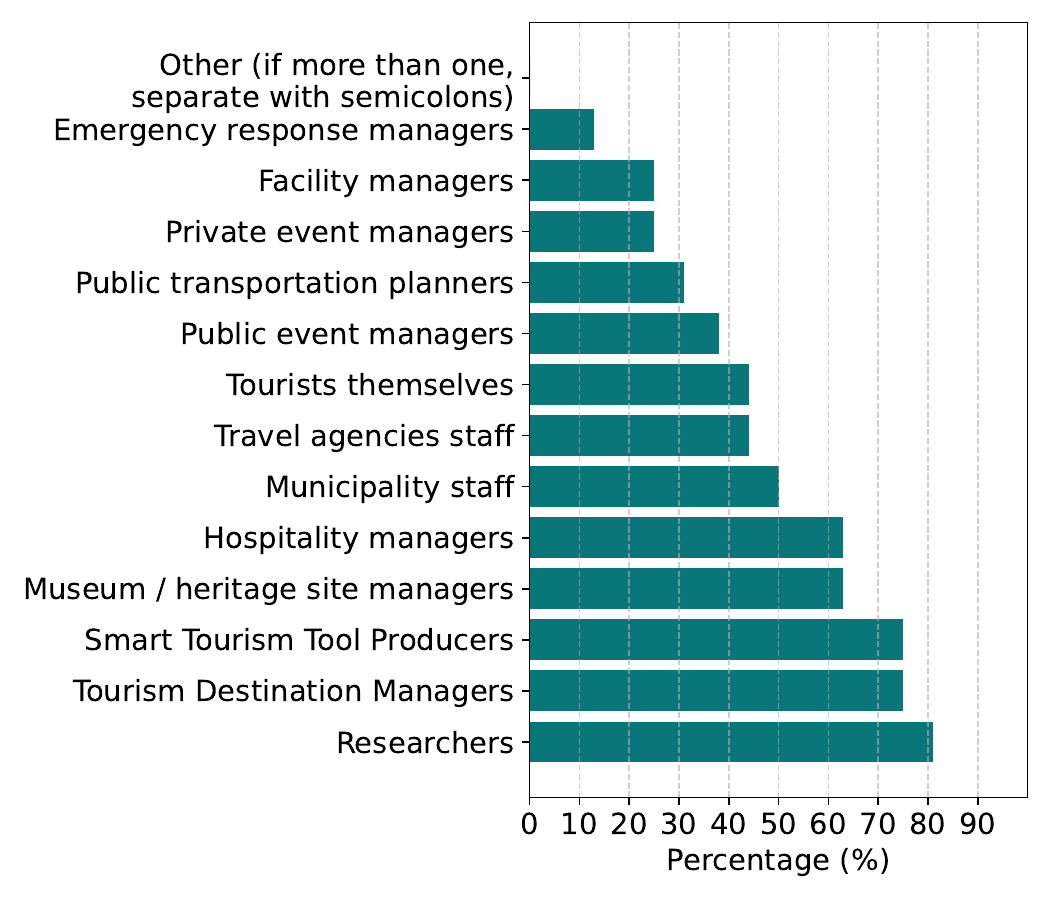}
        \caption{Type of professionals most likely to use the\\Observatory}
        \label{fig:types}
    \end{minipage}%
    \begin{minipage}[b]{0.35\textwidth}
        \centering
        \includegraphics[width=\textwidth]{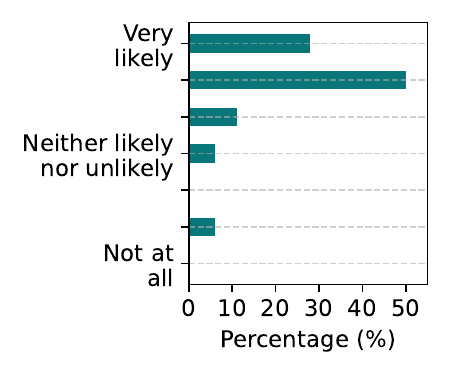}
        \caption{Likelihood of future use of the observatory}
        \label{fig:likelihood}
    \end{minipage}
\end{figure}

\begin{figure}[ht!]
    \centering
    \includegraphics[width=0.75\textwidth]{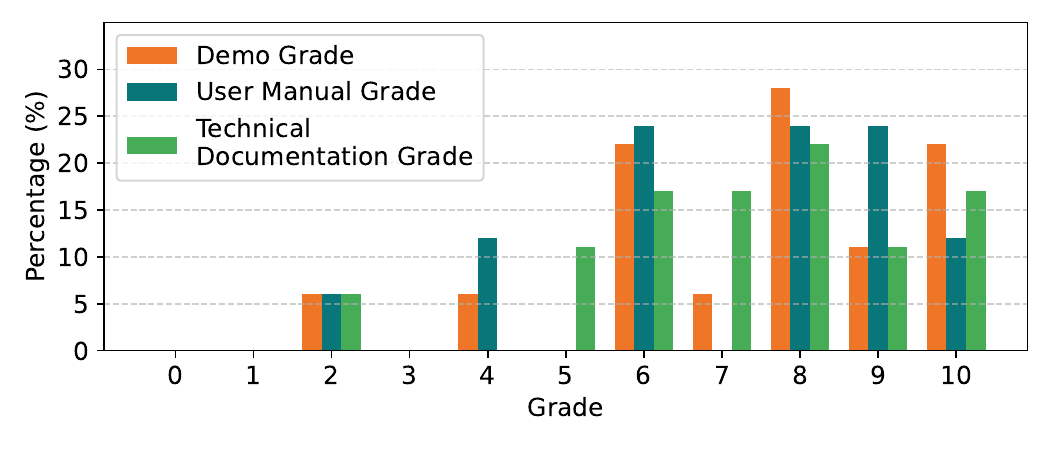}
    \caption{Grades of Demo, User Manual, and Technical Documentation}
    \label{fig:grades}
\end{figure}

The survey can be found at \href{https://tinyurl.com/sttobservatory-survey}{https://tinyurl.com/sttobservatory-survey}, and at the time of writing, it had already received 40 responses, most of them complete (not all questions were mandatory). Preliminary results show that researchers, producers of Smart Tourism Tools (STTs), and managers of tourism-related businesses are likely to find the Observatory most useful (Figure \ref{fig:types}). Tourists themselves and public service managers will also benefit. Figure \ref{fig:grades} shows the grades received for the demo, user manual, and technical documentation. We can see that they all received high scores (7 or above) in most responses, reflecting good, though not perfect, quality. In addition, most of the respondents said they were likely to use the Observatory in the future (Figure \ref{fig:likelihood}). 

Among the suggested functionalities, providing information on events and research publications stands out as a feasible addition to the European STT Observatory. Among the remaining suggestions, the most interesting were:

\begin{itemize}
    \item \textit{''Easy way to submit STTs''}
    
    \item \textit{''I recommend you include new technologies on your website, such as digital twins, cybersecurity, metaverse, generative AI, and quantum computing in smart tourist destinations. These disruptive tools are being used in some STDs (Smart Tourism Destinations), and they will change the new paradigm of travel and tourism activities in tourism cities and STDs.''}
    
\end{itemize}


\subsubsection{Monitoring}

\begin{figure}[ht!]
    \centering
    \includegraphics[width=0.9\textwidth]{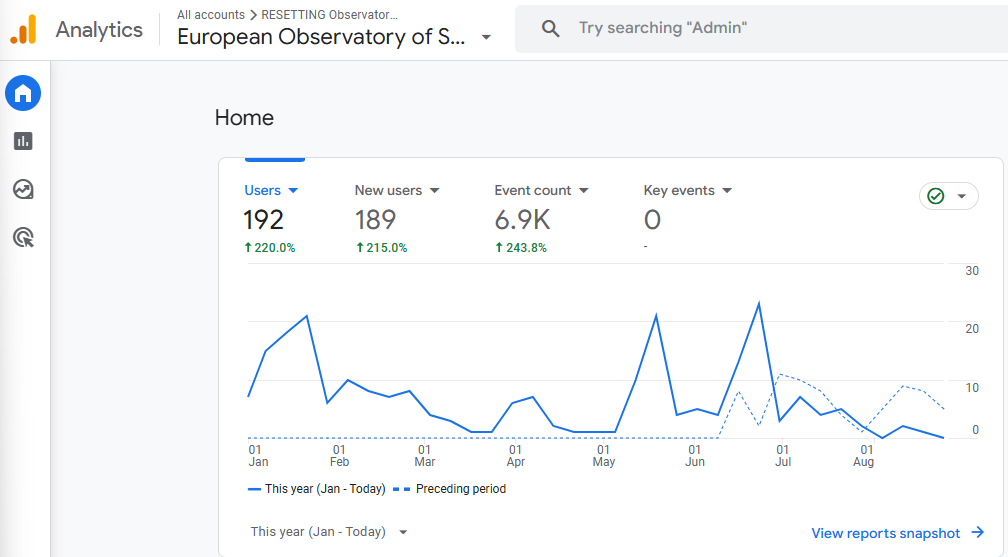}
    \caption{Monitoring of the Observatory with Google Analytics}
    \label{fig:google_analytics}
    \vspace{-0.15cm}
\end{figure}

The Google Analytics platform is a crucial tool used to monitor access to the Observatory. The snapshot below (Figure~\ref{fig:google_analytics}) corresponds to the last eight months (from January to August 2024). This robust monitoring process ensures the Observatory's performance is meticulously tracked and any necessary adjustments can be made to maintain its effectiveness.
 
\section{Conclusions}
\label{sec:conclusions}


The initial objective of this paper was to create a smart ETL process, where "smart" means integrating AI into a conventional ETL framework. The extraction phase was completed by extracting various types of content, including images, text, links, QR codes, phone numbers, email addresses, and more, from STT catalogs in PDF format. The transformation phase focused on developing an automatic classification system for STTs using AI, particularly LLMs. That classification (STT labeling) will facilitate tourism sector SMEs searching for innovative and sustainable business solutions.

It is essential to strive for high accuracy in classification, but achieving a perfect one (100\%) is unnecessary because eliminating all errors is often unfeasible. The results demonstrated that classification is feasible but lacked the consistency required for large-scale implementation. This could be due to either the use of potentially inappropriate few-shot examples or the volatility of the LLM’s reasoning capabilities, which, as an online model, can sometimes underperform due to overload. There is also the problem of hallucinations in LLMs. In this paper, the model occasionally classified an STT as if it understood the problem but then returned categories that were not part of the given STT taxonomy.
In some cases, it even used information from previously classified STTs to justify the classification of the current STT. A recent study, \cite{Banerjee2024}, examined the issue of hallucinations in LLMs and found that while there are techniques to mitigate them, these hallucinations are unavoidable, and systems using LLMs must be prepared to deal with them. The loading phase was completed by integrating the content into the European STT Observatory using the API of its hosting platform.

Although the results were not as optimal as desired, the few successful tests conducted in a volatile environment (online LLM) indicate that an acceptable classification accuracy could be achieved in a controlled environment (local LLM).



\section{Future work}
\label{sec:future_work}

Given the promising initial results for the LLM-based classification, there are several pathways for future work in STT's labeling, as follows:

\begin{itemize}
    \item Extract and classify content from the STTs' URLs found in catalogs. The main challenge here is that those pages do not have a standard layout as in the catalogs, so LLMs are also required for content extraction. If the content is then segmented, such as in catalogs, then the technique proposed in this paper can be used for STT classification (labeling).
    
    \item Identify new STTs by web searches automatically. The main challenge here is finding reliable STT candidates. Analyzing the outputs of a search engine with an LLM for soundness checking seems to be a viable first step. Then, a validation step performed by a human actor would allow applying a reinforcement learning approach to fine-tune the selection process. Once a sound STT candidate is found, we will resort to the previous pending problem.

    \item Allow STT producers to independently update the information on their STTs made available in the observatory. This will require hosting the open-source version of Omeka in a dedicated server and implementing a new \href{https://omeka.org/classic/docs/Admin/Adding_and_Managing_Plugins/}{Omeka plugin} for user access control, therefore extending Omeka's core functionality. By customizing the user roles and permission system, associating collections with specific users, and intercepting workflow processes, it is possible to create a flexible and secure access control system. That plugin will probably use design patterns like the Observer, Factory, or Strategy to enforce clean and maintainable access control logic.
   
\end{itemize}

Other research threads may also be explored beyond the observatory context:

\begin{itemize}
    \item Experimenting with open LLMs such as the ones made available in the \href{https://huggingface.co/models}{HuggingFace Hub}, in dedicated hardware, to mitigate the effect of user load in limited access commercial LLMs, such as Microsoft Copilot used herein. 
    
    \item Experimenting with fine-tuning an LLM using frameworks like PyTorch, TensorFlow, or Hugging Face’s transformers library. These provide the tools to manage and customize training procedures, such as learning rate schedules, gradient accumulation, etc.
    
    \item Experimenting with the Retrieval Augmented Generation (RAG) concept. RAG involves retrieving relevant information from an external knowledge base before the LLM generates a response. This ensures that the LLM only accesses the necessary information not present in its training data, resulting in more accurate and contextually relevant answers.
    
\end{itemize}


\section*{Acknowledgements}
This work was partially funded by the \href{https://www.fct.pt/en/}{Portuguese Foundation for Science and Technology (FCT)}, under \href{https://istar.iscte-iul.pt/}{ISTAR-Iscte} project UIDB/04466/2020 and \href{https://www.cense.fct.unl.pt/}{CENSE NOVA-FCT} project UIDB/04085/2020.

%
%
%
\bibliographystyle{splncs04}
\bibliography{MAIN}


\end{document}